\documentclass[12pt]{article}
\usepackage{amsfonts,pifont}
\usepackage{amsmath,chemarrow}
\usepackage{amssymb}
\usepackage{amstext}
\usepackage{amsfonts}
\usepackage{graphicx}
\usepackage{indentfirst}
\usepackage{hyperref,comment}
\makeatletter
\def\ExtendSymbol#1#2#3#4#5{\ext@arrow 0099{\arrowfill@#1#2#3}{#4}{#5}}
\def\RightExtendSymbol#1#2#3#4#5{\ext@arrow 0359{\arrowfill@#1#2#3}{#4}{#5}}
\def\LeftExtendSymbol#1#2#3#4#5{\ext@arrow 6095{\arrowfill@#1#2#3}{#4}{#5}}
\makeatother

\begin{document}
\baselineskip 20pt

\title{Estimating the mass of the hidden charm $1^+(1^{+})$ tetraquark state via QCD sum rules}
\author{Cong-Feng Qiao\footnote{qiaocf@ucas.ac.cn} and
Liang Tang\footnote{tangl@ucas.ac.cn}\\[0.5cm]
Department of Physics, University of Chinese Academy of Sciences \\
YuQuan Road 19A, Beijing 100049, China}

\date{}
\maketitle

\begin{abstract}

By using QCD sum rules, the mass of the hidden charm tetraquark
$[cu][\bar{c}\bar{d}]$ state with $I^{G} (J^{P}) = 1^+ (1^{+})$
(HCTV) is estimated, which presumably will turn out to be the newly observed
charmonium-like resonance $Z_c^+(3900)$. In the calculation,
contributions up to dimension eight in the operator product
expansion(OPE) are taken into account. We find $m_{1^+}^c =
(3912^{+306}_{-153}) \, \text{MeV}$, which is consistent, within the
errors, with the experimental observation of $Z_c^+(3900)$.
Extending to the b-quark sector, $m_{1^+}^b = (10561^{+395}_{-163})
\,\text{MeV}$ is obtained. The calculational result strongly supports the
tetraquark picture for the ``exotic'' states of $Z_c^+(3900)$ and
$Z_b^+(10610)$.

\end{abstract}

\section{Introduction}

Recently, the BESIII Collaboration reported the observation of a new
charged charmonium-like state in the $J/\psi \pi^\pm$ channel in
$Y(4260) \rightarrow J/\psi \pi^+ \pi^-$ decay
\cite{Ablikim:2013mio}. Its mass and width are $(3899.0 \pm 3.6 \pm
4.9) \, \text{MeV}$ and $(46 \pm 10 \pm 20) \, \text{MeV}$,
respectively. Soon afterwards, the Belle \cite{Liu:2013dau} and CLEO
\cite{Xiao:2013iha} Collaborations confirmed the existence of this
hadronic structure. Notice that this new resonance, nominated as
$Z_c^+(3900)$, is a charged charmonium-like state; therefore, it
certainly contains at least four quarks, a pair of charm quarks and two
light quarks. It is an exotic state. In the b-quark sector,
recall that two bottom-like charged sates $Z_b^+(10610)$ and
$Z_b^+(10650)$ were observed by the Belle Collaboration
\cite{Collaboration:2011gja, Belle:2011aa}. That implies that there exist
similar structures in the charm and bottom energy regions. These new
findings reflect the renaissance of the study of the so-called exotic states.

In the literature, various models have been proposed to interpret
the new experimental observations. For $Z_c^+(3900)$, for instance,
models of the molecular state \cite{Wang:2013cya, Cui:2013yva,
Zhang:2013aoa, Guo:2013sya, Ke:2013gia}, the tetraquark state
\cite{Faccini:2013lda, Dias:2013xfa, Braaten:2013boa}, the initial
single pion emission (ISPE) scheme \cite{Chen:2013coa} and so on were
proposed. For a comprehensive review of the theoretical status of
this state, we refer the reader to Ref.\cite{Voloshin:2013dpa}.
Since a definite conclusion has not yet been reached, more efforts are
still necessary to explore its inner structure.

The method of QCD sum rules \cite{Shifman, Reinders:1984sr,
Narison:1989aq, P.Col} has been applied successfully to many
hadronic phenomena, such as the hadron spectrum and hadron decays. In
this approach, an interpolating current with proper quantum numbers
are constructed corresponding to a hadron of interest. Then, by
constructing a correlation function and matching its operator
product expansion (OPE) to its hadronic saturation, the main
function for extracting the mass or decay rate of the hadron is
established. In the original paper on the quark model
\cite{GellMann:1964nj}, Gell-Mann discussed the possibility of the
existence of free diquarks. The concept of diquark is based on the
fundamental theory, and has been invoked to interpret a number of
phenomena observed in experiment \cite{Jaffe:2004ph, Wilczek:2004im,
Maiani:2007wz}. In Ref.\cite{Matheus:2006xi}, the exotic state
X(3872) was explored through the QCD sum rules, where the hadronic
state was considered as a hidden charm tetraquark state with quantum
number $I^G (J^{PC}) = 0^+ (1^{++})$ (HCTS). Employing the same
interpolating current, Chen and Zhu investigated the $1^{+-}$
tetraquark state and found its mass to be $(4.02 \pm 0.09) \,
\text{GeV}$ \cite{Chen:2010ze}.

In this paper, we calculate the mass of the hidden charm tetraquark
state with $I^G (J^P) = 1^+ (1^+)$ (HCTV) by using the QCD sum
rules, and confront it with the $Z_c^+(3900)$. Here, the HCTV is
interpreted as the isospin 1 partner of the HCTS. Comparing this
work with Ref.\cite{Matheus:2006xi}, two differences are noteworthy.
First, the interpolating current here is different from the HCTS
current. Second, of the HCTV, as mentioned in
Ref.\cite{Matheus:2006xi}, the higher-dimensional two-gluon and
mixed condensates are not negligible in order to obtain a reasonable
sum rule. Hence, in this work, the non-perturbative condensates up to
dimension eight are taken into account. In addition, different from
Refs.\cite{Matheus:2006xi, Chen:2010ze, Cui:2011fj} on HCTV, in our
analysis the quark-gluon condensate term in the light-quark ``full"
propagator is considered, and a moderate criterion is adopted in
finding the available threshold parameter $\sqrt{s_0}$ and the Borel
window $M_B^2$.

\section{Formalism}

The starting point of the QCD sum rules is the two-point correlation
function constructed from the interpolating current:
\begin{eqnarray}
\Pi_{\mu \nu}(q) = i \int d^4 x e^{i q \cdot x} \langle 0 | T
\big{\{} j_\mu(x) j^\dagger_\nu(0) \big{\}} | 0 \rangle \; .
\end{eqnarray}

The interpolating current of the HCTV is expressed as
\cite{Dias:2013xfa}:
\begin{eqnarray}
j_\mu(x) = \frac{i \epsilon_{a b c} \epsilon_{d e c}}{\sqrt{2}}
\big[ \left( u^T_a(x) C \gamma_5 c_b(x) \right) \left( \bar{d}_d
\gamma_\mu C \bar{c}_e^T \right) - \left( u^T_a(x) C \gamma_\mu
c_b(x) \right) \left( \bar{d}_d \gamma_5 C \bar{c}_e^T \right)
\big] \; , \label{current}
\end{eqnarray}
where $a$, $b$, $c$, $\cdots$, are color indices, and $C$
represents the charge conjugation matrix. Note that there is a minus sign
difference between the current given in Eq.(\ref{current}) and the
one in Ref.\cite{Matheus:2006xi}. Therefore, even under the SU(2)
symmetry the mass obtained for the HCTV differs from the HCTS, which
is what is to be analyzed in the following.

Generally, the two-point correlation function takes the following
Lorentz covariance form:
\begin{eqnarray}
\Pi_{\mu \nu}(q) = - \left(g_{\mu\nu} -\frac{q_\mu q_\nu}{q^2} \right) \Pi_1(q^2) +
\frac{q_\mu q_\nu}{q^2} \Pi_0(q^2) \; .
\end{eqnarray}
Because the axial vector current is not conserved, there are two
independent parts appearing in the correlation function, i.e.
$\Pi_1(q^2)$ and $\Pi_0(q^2)$, where the subscripts 1 and 0 denote
the quantum numbers of the spin 1 and 0, respectively.

On the phenomenological side, after separating the ground state
contribution from the pole term in $\Pi_1(q^2)$, the correlation
function is expressed as a dispersion integral over a physical
regime, i.e.,
\begin{eqnarray}
\Pi_1(q^2) & = & \frac{\lambda_{1^+}^{c \; 2}}{m_{1^+}^{c \; 2} -
q^2} + \frac{1}{\pi} \int_{s_0}^\infty d s \frac{\rho^h(s)}{s - q^2} \; .
\label{spectraldensity}
\end{eqnarray}
Here, $m_{1^+}^c$ represents the HCTV mass, $\rho^h(s)$ is the
spectral density representing the contributions of higher excited
and continuum states, $s_0$ denotes the threshold of higher excited
and continuum states, and $\lambda_{1^+}^c$ stands for the pole
residue, representing the coupling strength defined by $\langle 0 |
j_\mu |\text{HCTV} \rangle = \lambda_{1^+}^c \epsilon_\mu$.

On the OPE side of $\Pi_1(q^2)$, the correlation function can be
expressed as a dispersion relation:
\begin{eqnarray}
\Pi_{1}^{OPE}(q^2) = \int_{4 m_c^2}^{\infty} d s
\frac{\rho^{OPE}(s)}{s - q^2} + \Pi_1^{\langle g_s \bar{q} \sigma
\cdot G q \rangle \langle \bar{q} q \rangle}(q^2) + \Pi_1^{\langle
g_s^2 G^2 \rangle^2}(q^2) \; . \label{eq5}
\end{eqnarray}
Here, $\rho^{OPE}$ is given by the imaginary part of the correlation
function, $ \rho^{OPE}(s) = \text{Im} [\Pi_1^{OPE}(s)]/{\pi} $ and
it can be written as
\begin{eqnarray}
\rho^{OPE}(s) & = & \rho^{pert}(s) + \rho^{\langle \bar{q} q
\rangle}(s) + \rho^{\langle g_s^2 G^2 \rangle}(s) + \rho^{\langle
g_s \bar{q} \sigma \cdot G q \rangle}(s) + \rho^{\langle \bar{q} q
\rangle^2}(s) + \rho^{\langle g_s^3 G^3 \rangle}(s) \nonumber \\
& + & \rho^{\langle g_s \bar{q} \sigma \cdot G q \rangle \langle
\bar{q} q \rangle}(s) + \rho^{\langle g_s^2 G^2 \rangle^2} + \cdots
\; , \label{eq6}
\end{eqnarray}
where ``$\cdots$" stands for other higher dimension condensates
neglected in this work. $\Pi_1^{\langle g_s \bar{q} \sigma \cdot G q
\rangle \langle \bar{q} q \rangle}(q^2)$ and $\Pi_1^{\langle g_s^2
G^2 \rangle^2}(q^2)$ denote those contributions of the correlation
function which have no imaginary parts but have nontrivial values
under the Borel transform. After making the Borel transform on the
OPE side, we get
\begin{eqnarray}
\Pi_1^{OPE} (M_B^2) = \int_{4m_c^2}^{\infty} d s
\rho^{OPE}(s) e^{-s/M_B^2} + \Pi_1^{\langle g_s \bar{q}
\sigma \cdot G q \rangle \langle \bar{q} q \rangle} (M_B^2) +
\Pi_1^{\langle g_s^2 G^2 \rangle^2} (M_B^2) \; .
\end{eqnarray}

To evaluate the spectral density, the ``full" propagators $S^q_{i
j}(x)$ and $S^Q_{i j}(p)$ for light ($q=u$, $d$ or $s$) and heavy
quarks ($Q=c$ or $b$) are necessary, in which the vacuum condensates
are explicitly shown \cite{Reinders:1984sr}, i.e.,
\begin{eqnarray}
S^q_{i j}(x) \! \! \! & = & \! \! \! \frac{i \delta_{i j} \hat{x}}{2
\pi^2 x^4} - \frac{m_q \delta_{i j}}{4 \pi^2 x^2} - \frac{i g_s
t^a_{i j} G^a_{\kappa \lambda}} {32 \pi^2 x^2}(\sigma^{\kappa
\lambda} \hat{x} + \hat{x} \sigma^{\kappa \lambda}) + \frac{i
\delta_{i j} \hat{x}}{48} m_q \langle \bar{q} q \rangle -
\frac{\delta_{i j}
\langle \bar{q} q \rangle}{12} \nonumber \\
\! \! \! & - & \! \! \! \frac{\delta_{i j} \langle g_s \bar{q}
\sigma G q \rangle x^2}{192} - \frac{t^a_{i j} \sigma^{\kappa^\prime
\lambda^\prime}}{192} \langle
g_s \bar{q} \sigma \cdot G^\prime q \rangle + \cdots \;, \\
S^Q_{i j}(p) \! \! \! & = & \! \! \! \int \frac{d^4 p}{(2 \pi)^4}
e^{-i p \cdot x} \bigg\{ \frac{i}{\hat{p} - m_Q} \delta_{i j} -
\frac{i}{4} g_s (t^c)_{i j} G^c_{\kappa \lambda} \frac{1}{(p^2 -
m_Q^2)^2}
\nonumber \\
\! \! \! & \times & \! \! \! [\sigma^{\kappa \lambda} (\hat{p} +
m_Q) + (\hat{p} + m_Q) \sigma^{\kappa \lambda}] + \frac{i}{12} g_s^2
\delta_{i j} G^a_{\alpha \beta} G^a_{\alpha \beta} m_Q \frac{p^2 +
m_Q \hat{p}}{(p^2 - m_Q^2)^4}
\nonumber \\
\! \! \! & + & \! \! \! \frac{i \delta_{i j}}{48} \bigg[
\frac{(\hat{p} + m_Q) [\hat{p} (p^2 - 3 m_Q^2) + 2 m_Q (2 p^2 -
m_Q^2)] (\hat{p} + m_Q)}{(p^2 - m_Q^2)^6} \bigg] \langle g_s^3 G^3
\rangle + \cdots \bigg\} \; .
\end{eqnarray}
Here, $G^\prime$ represents the outer gluon field and the Lorentz
indices $\kappa^\prime$ and $\lambda^\prime$ are indices of the
outer gluon field coming from another propagator
\cite{Albuquerque:2013ija}.

We calculate the spectral density $\rho^{OPE}(s)$ up to dimension
eight at the leading order in $\alpha_s$ by the standard technique
of QCD sum rules. In order to find the difference between HCTV and
HCTS, we keep not only terms linear in the light-quark masses $m_u$
and $m_d$, but also the two-gluon and the quark-gluon mixed
condensates up to dimension eight. Through a lengthy calculation,
the spectral densities on the OPE side are obtained as:
\begin{eqnarray}
\rho^{pert}(s) \! \! \! & = & \! \! \! \frac{1}{2^{10} \pi^6}
\int^{\alpha_{max}}_{\alpha_{min}} \frac{d \alpha}{\alpha^3}
\int^{1 - \alpha}_{\beta_{min}} \frac{d \beta} {\beta^3} (1 -
\alpha - \beta) (1 + \alpha + \beta) {\cal F} (\alpha, \beta, s)^4
\nonumber \\[2mm]
\! \! \! & + & \! \! \! \frac{(m_u + m_d) m_c}{2^9 \pi^6}
\int^{\alpha_{max}}_{\alpha_{min}} \frac{ d \alpha}{\alpha^2}
\int^{1 - \alpha}_{\beta_{min}} \frac{d \beta}{\beta^3}(\alpha +
\beta - 1) (3 + \alpha + \beta) {\cal F} (\alpha, \beta, s)^3 \; ,
\label{rhopert} \\[2mm]
\rho^{\langle \bar{q} q \rangle}(s) \! \! \! & = & \! \! \! -
\frac{m_c \langle \bar{q} q \rangle} {2^5 \pi^4}
\int^{\alpha_{max}}_{\alpha_{min}} \frac{d \alpha}{\alpha^2} \int^{1
- \alpha}_{\beta_{min}} \frac{d \beta}{\beta} (1 + \alpha + \beta)
{\cal F}
(\alpha, \beta, s)^2 \nonumber \\[2mm]
\! \! \! & + & \! \! \! \frac{(m_u + m_d) \langle \bar{q} q
\rangle}{2^6 \pi^4} \bigg[ \int_{\alpha_{min}}^{\alpha_{max}}
\frac{d \alpha}{\alpha(1 - \alpha)} {\cal H} (\alpha, s)^2 -
\int^{\alpha_{max}}_{\alpha_{min}} \frac{d \alpha}{\alpha} \int^{1 -
\alpha}_{\beta_{min}} \frac{d \beta}{\beta} {\cal F}
(\alpha, \beta, s)^2 \nonumber \\[2mm]
\! \! \! & + & \! \! \! 4 m_c^2 \int^{\alpha_{max}}_{\alpha_{min}}
\frac{d \alpha}{\alpha} \int^{1 - \alpha}_{\beta_{min}} \frac{d
\beta}{\beta} {\cal F} (\alpha, \beta, s) \bigg] \; , \\[2mm]
\rho^{\langle g_s^2 G^2 \rangle}(s) \! \! \! & = & \! \! \!
\frac{\langle g_s^2 G^2 \rangle}{3 \times 2^9 \pi^6}
\int^{\alpha_{max}}_{\alpha_{min}} \! \! \! d \alpha \int^{1 -
\alpha}_{\beta_{min}} \! \! \frac{d \beta}{\beta^2} \bigg[
\frac{m_c^2 (1 - (\alpha + \beta)^2)}{\beta} - \frac{(1 - 2 \alpha - 2
\beta)}{2 \alpha} {\cal F} (\alpha, \beta, s)\bigg] \nonumber \\[2mm]
\! \! \! & \times & \! \! \! {\cal F} (\alpha, \beta, s)  -
\frac{m_c^2 \langle g_s^2 G^2 \rangle}{3 \times 2^{12} \pi^6}
\int^{\alpha_{max}}_{\alpha_{min}} \frac{d \alpha}{\alpha} \int^{1 -
\alpha}_{\beta_{min}} \frac{d \beta}{\beta} \bigg[ (1 + \alpha +
\beta) \nonumber \\[2mm]
\! \! \! &\times& \! \! \! - \frac{(1 - \alpha - \beta)(\alpha +
\beta + 3)}{\alpha} + \frac{1}{4 \alpha \beta} (\alpha + \beta - 1)^2
(\alpha + \beta + 5) \bigg] {\cal F} (\alpha, \beta, s) \; ,
\label{O4} \\[2mm]
\rho^{\langle g_s \bar{q} \sigma \cdot G q \rangle}(s) \! \! \! & =
& \! \! \! - \frac{m_c \langle g_s \bar{q} \sigma \cdot G q
\rangle}{2^6 \pi^4} \bigg[ 2 \int^{\alpha_{max}}_{\alpha_{min}}
\frac{d \alpha}{\alpha} {\cal H} (\alpha, s) -
\int^{\alpha_{max}}_{\alpha_{min}} \! \! \! d \alpha \int^{1 -
\alpha}_{\beta_{min}} \! \! \! d \beta \left( \frac{1}{\alpha} +
\frac{\alpha + \beta}{\beta^2} \right) \nonumber \\[2mm]
\! \! \! &\times& \! \! \! {\cal F} (\alpha, \beta, s) \bigg] +
\frac{m_c \langle g_s \bar{q} \sigma \cdot G q \rangle}
{3 \times 2^8 \pi^4} \int^{\alpha_{max}}_{\alpha_{min}}
\frac{d \alpha}{\alpha}\bigg[2 {\cal H}(\alpha, s) \nonumber \\[2mm]
\! \! \! & - & \! \! \! \int^{1 - \alpha}_{\beta_{min}} d
\beta \left( 1 + \frac{(1 + \alpha + \beta)}{\alpha} \right)
{\cal F} (\alpha, \beta, s) \bigg] \; , \label{O5} \\[2mm]
\rho^{\langle \bar{q} q \rangle^2}(s) \! \! \! & = & \! \! \!
\frac{\langle \bar{q} q \rangle^2}{12 \pi^2} m_c^2 \sqrt{1 - 4 m_c^2
/ s} \; , \label{O6}
\end{eqnarray}
\begin{eqnarray}
\rho^{\langle g_s^3 G^3 \rangle}(s) \! \! \! & = & \! \! \!
\frac{\langle g_s^3 G^3 \rangle}{3 \times 2^{10} \pi^6}
\int^{\alpha_{max}}_{\alpha_{min}} d \alpha \int^{1 -
\alpha}_{\beta_{min}} \frac{d \beta}{\beta^3} (1 - \alpha - \beta)
(1 + \alpha + \beta) \nonumber \\[2mm]
\! \! \! & \times & \! \! \! \bigg[m_c^2 \alpha + \frac{ {\cal F}
(\alpha, \beta, s)}{2} \bigg] \; , \label{O6G} \\[2mm]
\rho^{\langle g_s \bar{q} \sigma \cdot G q \rangle \langle \bar{q} q
\rangle}(s) \! \! \! & = & \! \! \! - \frac{\langle g_s \bar{q}
\sigma \cdot G q \rangle \langle \bar{q} q \rangle}{3^2 \times
2^5\pi^2} \int^{\alpha_{max}}_{\alpha_{min}}\alpha d \alpha \; ,
\label{rhoO8add} \\[2mm]
\rho^{\langle g_s^2 G^2 \rangle^2}(s) \! \! \! & = & \! \! \! -
\frac{23 \langle g_s^2 G^2 \rangle^2}{3^3 \times 2^{16}\pi^6}
\bigg[\int^{\alpha_{max}}_{\alpha_{min}} d \alpha \int^{1 -
\alpha}_{\beta_{min}} d \beta +
\int^{\alpha_{max}}_{\alpha_{min}} d \alpha \bigg] \; ,
\end{eqnarray}
and
\begin{eqnarray}
\Pi_1^{\langle g_s \bar{q} \sigma \cdot G q \rangle \langle
\bar{q} q \rangle} (M_B^2) \! \! \! & = & \! \! \! -
\frac{m_c^2 \langle g_s \bar{q} \sigma \cdot G q \rangle
\langle \bar{q} q \rangle}{24 \pi^2}\int^1_0 d \alpha
\bigg[ 1 + \frac{m_c^2}{\alpha (1 - \alpha) M_B^2} \nonumber \\
\! \! \! & - & \! \! \! \frac{5}{12 (1 - \alpha)} \bigg]
e^{- \frac{m_c^2}{\alpha (1 - \alpha) M_B^2}} \; , \; \; \;
\; \; \; \; \; \; \; \label{rhoO8} \\
\Pi_1^{\langle g_s^2 G^2 \rangle^2}(M_B^2) \! \! \! & = & \! \! \! -
\frac{11 m_c^2 \langle g_s^2 G^2\rangle^2}{3^2 \times 2^{18}\pi^6}
\int_0^1 d \alpha \int_0^{1 - \alpha} d \beta \frac{(1 - \alpha -
\beta)}{\alpha \beta} e^{-\frac{(\alpha + \beta) m_c^2}
{\alpha \beta M_B^2}} \nonumber \\
\! \! \! & + & \! \! \! \frac{m_c^4 \langle g_s^2 G^2
\rangle^2}{3^3 \times 2^{14} \pi^6} \left(1 + \frac{1}{M_B^2}
\right) \int_0^1 \frac{d \alpha}{\alpha^3} \int_0^{1 - \alpha}
\frac{d \beta} {\beta^3} \bigg[- \frac{(\alpha^2 + \beta^2)(1 -
\alpha - \beta)^2}{4} \nonumber \\
\! \! \! & - & \! \! \! \frac{(\alpha^3 + \beta^3)}{2} + (1 -
\alpha -\beta)(\alpha^3 + 2\alpha^2 + 2\beta^2 + \beta^3)\bigg]
e^{- \frac{(\alpha + \beta) m_c^2}{\alpha \beta M_B^2}} \; .
\label{eq19}
\end{eqnarray}
Here, $M_B$ is the Borel parameter introduced by the Borel
transform; we have the functions ${\cal F} (\alpha, \beta, s) = (\alpha +
\beta) m_c^2 - \alpha \beta s$ and ${\cal H} (\alpha, s) = m_c^2 -
\alpha (1 - \alpha) s$; the integration bounds are $\alpha_{min} =
(1 - \sqrt{1 - 4 m_c^2/s}) / 2$, $\alpha_{max} = (1 + \sqrt{1 - 4
m_c^2 / s}) / 2$ and $\beta_{min} = \alpha m_c^2 /(s \alpha -
m_c^2)$.

Matching the OPE side expression of the correlation function
$\Pi_1(q^2)$ with the phenomenological side one, and performing the Borel transform, one obtains a sum rule
for the corresponding HCTV mass. It reads
\begin{eqnarray}
m_{1^+}^{c}(s_0, M_B^2) = \sqrt{- \frac{R_1(s_0, M_B^2)}{R_0(s_0,
M_B^2)}} \; \label{mass-Eq}
\end{eqnarray}
with
\begin{eqnarray}
R_0(s_0, M_B^2) & = & \int_{4 m_c^2}^{s_0} d s \; \rho^{OPE}(s)
e^{- s / M_B^2} + \Pi_1^{\langle g_s \bar{q} \sigma \cdot G q
\rangle \langle \bar{q} q \rangle} (M_B^2) + \Pi_1^{\langle g_s^2
G^2 \rangle^2} (M_B^2) \; , \\
R_1(s_0, M_B^2) & = & \frac{\partial}{\partial{M_B^{-2}}}
{R_0(s_0, M_B^2)} \; .
\end{eqnarray}

It should be mentioned that in principle the four-gluon operator,
the $\langle g_s^2 G^2 \rangle^2$, also belongs to the
dimension-eight condensate, however, in practice we find it is only
1 \% of the mixed condensate $\langle g_s \bar{q} \sigma \cdot G q
\rangle \langle \bar{q} q \rangle$ in magnitude, and hence the four-gluon condensate is neglected in the evaluation of this work.
Moreover, in order to obtain a relatively reliable result through
the leading order calculation, one needs to depress the higher order
QCD corrections and hence to express the $m_{1^+}^{c}$ in terms of
Eq.(\ref{mass-Eq}), which is found to be less sensitive to the
radiative corrections than to the individual moments
\cite{Matheus:2006xi}.

\section{Numerical Analysis}

In performing the numerical evaluation, the values of the input
parameters, the condensates, and the quark masses are adopted as
follows \cite{Matheus:2006xi, Cui:2011fj, Narison:2002pw,SNB}:
\begin{eqnarray}
\begin{aligned}
& m_u = 2.3 \; \text{MeV} \; , & & m_d = 6.4 \; \text{MeV} \; , \\
& m_c (m_c) = (1.23 \pm 0.05) \; \text{GeV} \; , & & m_b (m_b) =
(4.24 \pm 0.06) \; \text{GeV}, \\
& \langle \bar{q} q \rangle = - (0.23 \pm 0.03)^3 \; \text{GeV}^3 \; ,
& & \langle g_s^2 G^2 \rangle = 0.88 \; \text{GeV}^4 \; , \\
& \langle \bar{q} g_s \sigma \cdot G q \rangle = m_0^2 \langle
\bar{q} q \rangle \; , & & \langle g_s^3 G^3 \rangle = 0.045 \;
\text{GeV}^6 \; ,\\
& m_0^2 = 0.8 \; \text{GeV}^2\; . & &
\end{aligned}
\end{eqnarray}
Here, the scale dependence of these parameters is not taken into
account since our calculation is performed at the leading order in
$\alpha_s$. The quark masses used here are evaluated in
Ref. \cite{SNB} by virtue of the QCD sum rules and hence they are
defined in the $\overline{\rm MS}$-scheme. For more details of the
nature of the inputs, one may refer to Ref.\cite{Matheus:2006xi}.

\begin{figure}
\begin{center}
\includegraphics[width=7.5cm]{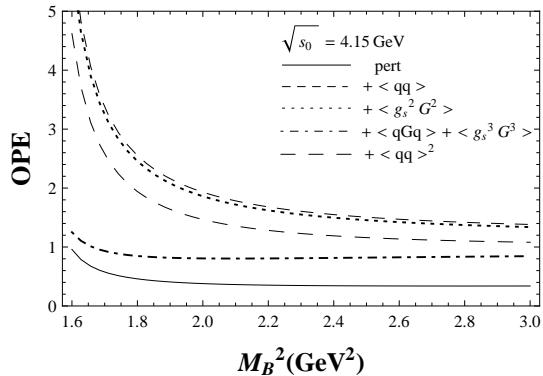}
\caption{The OPE convergence in the region $1.6 \leq M_B^2 \leq 3.0
\, \text{GeV}^2$ at $\sqrt{s_0} = 4.15 \, \text{GeV}$. The solid
line denotes the fraction of the perturbative contribution, and each
subsequent line denotes the addition of one extra condensate
dimension in the expansion, {\it i.e.}, $ \langle \bar{q} q \rangle$
(short-dashed line), $ \langle g_s^2 G^2 \rangle$ (dotted line),
$\langle g_s \bar{q} \sigma \cdot G q \rangle$ (dotted-dashed line),
and $\langle \bar{q} q \rangle^2$ (long-dashed line)}
\label{OPEconvergence}
\end{center}
\end{figure}

In the approach of QCD sum rules, choosing a proper threshold $s_0$
and Borel parameter $M_B^2$ are critical to obtain a reasonable
result. There are two criteria in making such choices \cite{Shifman,
Reinders:1984sr, P.Col}. First, the convergence of the OPE should be
kept. To this aim, one may compare the relative contribution of each
term in Eqs.(\ref{rhopert}) to (\ref{eq19}) with the total
contribution on the OPE side, which are shown in
Fig.\ref{OPEconvergence}. From the figure, we notice that a quite
good OPE convergence occurs when $M_B^2 \geq 1.9 \; \text{GeV}^2$,
and then we fix the lower working limit for $M_B^2$.

\begin{figure}
\begin{center}
\includegraphics[width=7.5cm]{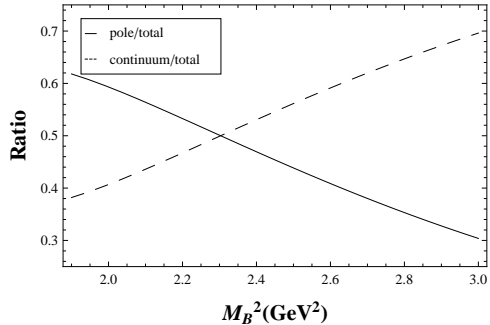}
\caption{The relative pole and continuum contributions at
$\sqrt{s_0} = 4.15 \; \text{GeV}$. The solid line represents
the relative pole contribution, and the dashed line corresponds
to the relative continuum contribution} \label{polecontribution}
\end{center}
\end{figure}

The second criterion to constrain the $M_B^2$ is that the pole
contribution should be larger than the continuum contribution. That
means we need to evaluate the relative pole contribution (PC) to the
total, the pole plus continuum, for various values of $M_B^2$. To
eliminate the contributions from the higher excited and continuum states
properly, we ask the pole contribution to be larger than $50\%$
\cite{P.Col, Matheus:2006xi}, which is a little different from the
constraint in \cite{Chen:2010ze}. The relative weight is presented
in Fig.\ref{polecontribution}, which tells the upper limit for
$M_B^2$. We note that the upper constraint on $M_B^2$ depends on the
threshold value $s_0$. So, for different $s_0$, we will find
different upper bounds for $M_B^2$. To determine the proper value of
$s_0$, we carry out a similar analysis to Ref.\cite{Matheus:2006xi}, and find that the optimal value of $s_0$
obtained there is also suitable in our case. The reason is that the
dominant contributions of the OPE side are the same in this work and
Ref.\cite{Matheus:2006xi}. Thus, for the proper $s_0$ in our
analysis, $\sqrt{s_0} = 4.15 \; \text{GeV}$, we find $M_B^2 \leq 2.3
\text{GeV}^2$.

\begin{figure}
\begin{center}
\includegraphics[width=7.5cm]{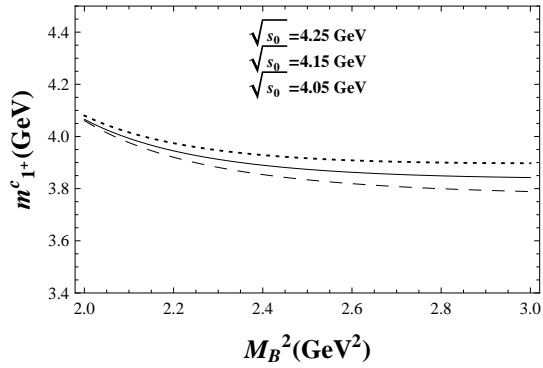}
\caption{The dependence of $m_{1^+}^c$ on the Borel parameter $M_B^2$,
where the three lines from bottom to top correspond to the continuum
threshold $\sqrt{s_0}$ being $4.05$, $4.15$, $4.25 \, \text{GeV}$,
respectively} \label{mass}
\end{center}
\end{figure}

Since the interpolating current in Eq.(\ref{current}) is different
from what in Ref.\cite{Matheus:2006xi}, the OPE contributions in
this work and in the HCTS analysis must be different. To highlight the
contributions of the new high-dimensional condensates in the HCTV,
in Table.\ref{ratio} we present the relative ratios of the
additional terms to the existing terms in Ref.\cite{Matheus:2006xi}
for each involved condensate at $\sqrt{s_0} = 4.15 \; \text{GeV}$.
Among these ratios in Table.\ref{ratio}, we find that the additional
contributions of dimension-four and -eight condensates are
considerable for the HCTV, which is different from the case in
Ref.\cite{Matheus:2006xi}. That is to say, the inclusion of high-dimensional condensates is necessary in obtaining a precise and
reliable mass of the HCTV. Fig.\ref{mass} shows the dependence
of $m_{1^+}^c$ on the Borel parameter $M_B^2$, where lines from bottom
to top correspond to the continuum threshold $\sqrt{s_0}$ being
$4.05$, $4.15$, $4.25 \, \text{GeV}$, respectively.

\begin{table}[h]
\caption{The relative ratios of the additional terms to those terms
in Ref.\cite{Matheus:2006xi} at $\sqrt{s_0} = 4.15 \; \text{GeV}$.
The subscripts denote the condensate dimensions. The
``$\text{ratio}_{O_4}$" denotes the ratio of the second term to the
first term in Eq.(\ref{O4}); the ``$\text{ratio}_{O_5}$" denotes the
second term to the first term in Eq.(\ref{O5});
``$\text{ratio}_{O_6}$" for Eq.(\ref{O6G}) to Eq.(\ref{O6}); and
``$\text{ratio}_{O_8}$" for Eq.(\ref{rhoO8add}) to Eq.(\ref{rhoO8}),
respectively}
\begin{center}
\begin{tabular}{|c|c|c|c|c|c|}
\hline\hline $M_B^2(\text{GeV}^2)$ & 1.6 & 1.9 & 2.2 & 2.5 & 2.8 \\
\hline $\text{ratio}_{O_4}$ & 0.45 & 0.39 & 0.34 & 0.31 & 0.29 \\
\hline $\text{ratio}_{O_5}$ & 0.03 & 0.03 & 0.03 & 0.04 & 0.04 \\
\hline $\text{ratio}_{O_6}$ & $< 0.01$ & 0.01 & 0.01 & 0.01 & 0.01
\\ \hline $\text{ratio}_{O_8}$ & 0.06 & 0.07 & 0.08 & 0.10 & 0.11 \\
\hline \hline
\end{tabular}
\end{center}
\label{ratio}
\end{table}

In the end, we obtain the HCTV mass:
\begin{eqnarray}
m_{1^+}^c = (3912^{+306}_{-153}) \, \text{MeV} \; .
\end{eqnarray}
Here, the errors stem from the uncertainties of the Borel parameter
$M_B$, the charm quark mass, the condensates, and the threshold parameter $s_0$. Note that the difference between the upper error and the lower error is due to the mass asymmetry in the Borel window.

\section{Summary and Conclusions}

In the approach of QCD sum rules, hadrons are represented by their interpolating quark currents taken with large virtualities. In this work, in order to extract the mass of the HCTV, we have constructed the proper interpolating current with the quantum numbers of $I^G (J^P) = 1^+ (1^+)$, which coincide with the newly observed charged charmonium-like resonance $Z_c^+(3900)$.

In our calculation, the non-perturbative QCD contributions up
to dimension eight in the OPE are taken into account. We find that the
$1^+$ hidden charm tetraquark state lies in around 3900 MeV, i.e.
$m_{1^+}^c = (3912^{+306}_{-153}) \, \text{MeV}$, which hence
presumably will turn out to be the newly observed charmonium-like resonance
$Z_c^+(3900)$. Comparing to a similar work of
Ref.\cite{Chen:2010ze}, where the mass of the hidden charm $1^{+-}$
tetraquark state with the same interpolating current under the
isospin symmetry was evaluated, we add a new mixed condensate term
in the light-quark propagator, which affects the contributions of
the dimension five and dimension eight in the OPE. Moreover, in
order to highlight the contribution of the ground state in
Eq.(\ref{spectraldensity}), in our analysis two constraint criteria
are employed.

We straightforwardly extend our analysis to the b-quark sector. With
the same quantum numbers, the mass of the hidden bottom tetraquark
state $[bu][\bar{b}\bar{d}]$ is obtained, i.e. $m_{1^+}^b =
(10561^{+395}_{-163}) \, \text{MeV}$ with $\sqrt{s_0} = 11.30 \,
\text{GeV}$ and $M_B^2 = 9.8 \, \text{GeV}^2$. This state has been
investigated via QCD sum rules in Ref.\cite{Cui:2011fj}, where only
the operators up to dimension six in OPE were considered and hence
the result is somehow different from ours. In our analysis,
operators of dimension eight are also taken into account. Our
calculational result, within uncertainties, strongly supports the
tetraquark picture of the state $Z_b^+(10610)$ observed in experiment
\cite{Collaboration:2011gja, Belle:2011aa}.

Finally, it should be mentioned that in order to make a
more solid prediction for the multiquark states in QCD sum rules,
the radiative correction and the energy-scale dependence on quark
masses and condensates in the calculation should be taken into
account, which are mostly missing in present-day investigations.

\vspace{.7cm} {\bf Acknowledgments} \vspace{.3cm}

This work was supported in part by the National Natural Science
Foundation of China(NSFC) under the Grants 10935012, 10821063,
11175249, and 11375200.


\end{document}